\documentclass{webofc}
\usepackage[varg]{txfonts}   
\begin{document}
\title{A Scientific Workflow System for Satellite Data Processing\\ with Real-Time Monitoring}
%
%

\author{
	\firstname{Minh Duc}
	\lastname{Nguyen}
	\inst{1}
	\thanks
	{\email{nguyendmitri@gmail.com}}
}

\institute{Skobeltsyn Institute of Nuclear Physics, Lomonosov Moscow State University, Moscow, Russia}

\abstract{%
  This paper provides a case study on satellite data processing, storage, and distribution in the space weather domain by introducing the Satellite Data Downloading System (SDDS). The approach proposed in this paper was evaluated through real-world scenarios and addresses the challenges related to the specific field. Although SDDS is used for satellite data processing, it can potentially be adapted to a wide range of data processing scenarios in other fields of physics.
}
\maketitle
\section{Introduction}
\label{intro}

The ultimate goal of the research in space weather is to create complex operational magnetosphere-ionosphere models which allow predicting the radiation risk for space satellites in different orbits, and also estimating the occurrence of risks of technological disasters due to magnetic storms and charged particle precipitation. The models use space-related data sets acquired by multiple satellites of various purposes and capabilities as well as ground stations to produce forecasts. Such models require all data sets to be collected at one place and also to be converted into a unified data format used by them as the input. Such task is challenging and time-consuming because data formats, storage methods, datasets and values of registered factors differ between satellites. The size of data sets can be huge so that the disk capacity of a personal computer might be insufficient for data storage. There is a high demand for creating an automatic system for collecting data from various satellites, processing them, converting them to a unified canonical format and making them available in a standardized machine-readable format.

Currently, there are several existing solutions to the problem such as SWX \cite{smdc}, xv\_SAT \cite{xv_sat, pthread_xv_sat}, the Herschel Data Processing System \cite{herschel}, and NJOY \cite{njoy}. The Satellite Data Downloading System (SDDS) presented in this paper has several advantages in comparison with the systems listed above. Firstly, SDDS does not act as a specific data processing system that targets a particular data source and has just one workflow. SDDS is a framework that eases the task of creating such systems that can have different workflows in data processing. Secondly, SDDS provides a web interface for data scientists, a RESTful API and a Python library for developers so they can build their systems upon SDDS. Thirdly, from the first day, SDDS has been developed as a modular system that can scale easily. Components of SDDS can run either on the same machine or different machines connected via a local network or the Internet. Different versions of a component can run interchangeably. Each component can be replaced with a new one just by editing the proper configuration file as long as the new component follows the convention. And finally, SDDS has a built-in real-time monitoring subsystem that helps operators to identify the scope of an error quickly and prevent it in the future. To our knowledge, SDDS is the first framework dedicated to satellite data processing that provides API for developers.

\section{Overview}
\label{sdds_overview}
SDDS has been designed to implement the whole workflow of a satellite data processing cycle. A cycle consists of six steps:
1) Connecting to data sources. A data source is a server where data of a satellite are collected and can be downloaded; 2) Downloading actual data (so-called raw data). Raw data of a satellite can be a big one file in which data of all instruments installed on the satellite are packed or a set of files each of which contains data of an instrument; 3) Extracting binary (level-0) data of each instrument from the raw data; 4) Processing level-0 data and producing scientific (level-1) data that are ready to be
loaded into a database; 5) Loading level-1 data to a database; 6) Compressing and moving raw, level-0, and level-1 data to storage.

\begin{figure}
\centering
\includegraphics[height=6.5cm, clip]{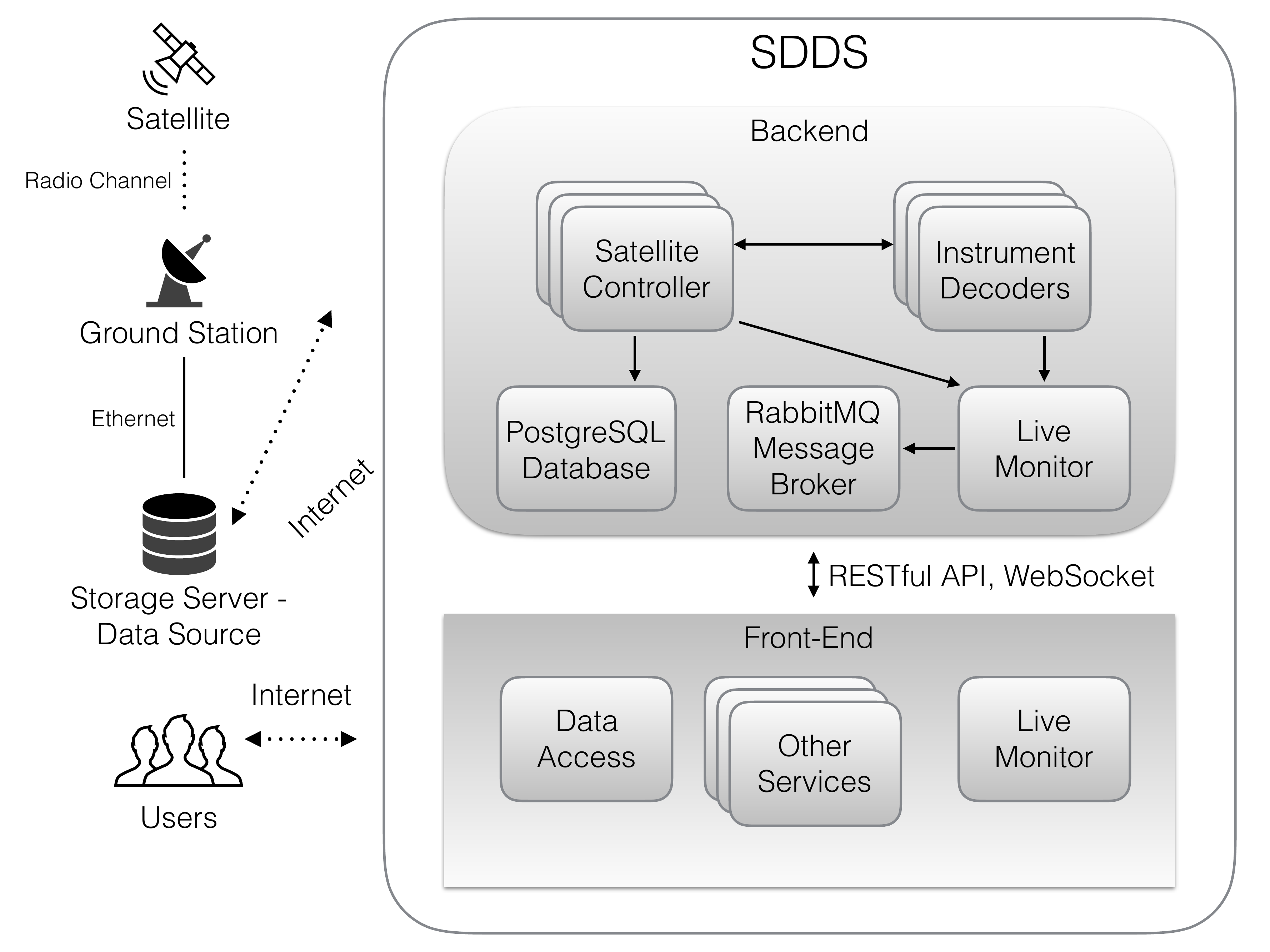}
\caption{The architectural overview of SDDS}
\label{fig-1}
\end{figure}

SDDS consists of two parts: a backend and a frontend. A detailed view of the architecture is shown in figure~\ref{fig-1}. Backend components have been developed using the Django Framework written in Python. Frontend components have been developed using the React library created by Facebook, Javascript, HTML, and CSS.

The backend part consists of the following components:
\begin{itemize}
\item A set of satellite controllers, each of which is responsible for a satellite and implement the whole data processing cycle. Each satellite controller uses decoders provided by satellite developers to decode level-0 data and produce level-1 data;
\item The Live Monitor subsystem which is responsible for monitoring all activities of all components in real time. The Live Monitor subsystem records each event during the data processing cycle and sends it to the RabbitMQ message broker;
\item The DBIface interface which is responsible for database interaction. The DBIface interface supports the following engines: PostgreSQL, MySQL, and Oracle;
\item The RabbitMQ message broker which is responsible for broadcasting all recorded events (short text messages) to all subscribed clients. A subscribed client is an instance of the Live Monitor's web interface that has established a WebSocket connection to the RabbitMQ message broker. The messages are sent using the STORM protocol;
\item A web server, which could be either Apache or Nginx, to serve RESTful requests.
\end{itemize}

On the frontend, there are two base components available currently:
\begin{itemize}
\item The Live Monitor's web interface shows recorded events (or current conditions) during the data processing cycle;
\item The Data Access services allow users to acquire data from the database. Data can be retrieved either by using the web interface or the RESTful API.
\end{itemize}

Each component of SDDS is a micro service with a RESTful API. Components can work together locally on the same computer or can be distributed across different machines connected via a local network or the Internet. Almost every component of SDDS has been developed as a library. Hence it gives users of SDDS the possibility to customize certain components to meet their requirements. It is also possible to change the behavior of a component by editing its configuration file which is written in the JSON format.

\section{Workflow automation}
\label{workflow}

The main component of SDDS is the satellite controller which is purely a Python class called SDDownloader. Each satellite controller in SDDS is dedicated for processing data of a certain satellite. Each step of the data processing cycle is implemented as a function of the SDDownloader class. Each function, in turn, is a composition of smaller functions (routines). Using the object oriented programming principles, one can create a customized class which is inherited from the SDDownloader class and redefine certain functions to meet their requirements.

Usually, data of a satellite are duplicated in several data sources. In this case, it is enough to download and process data from the first data source. But in many cases, data are distributed across data sources, and they need to be merged after processing. The SDDownloader class can handle all described scenarios. If a data source is temporarily or permanently unavailable, the satellite controller will retry and continue downloading data when the data source becomes available.

Different data sources have different security constraints and thus use different types of connection. At the moment the following connection types are supported by SDDS out-of-box: HTTP(S), FTP(S), PPTP, IPsec, HTTP(S)/FTP(S) over PPTP or IPsec. To support a new connection type one just needs to implement a function for this purpose and pass it as a parameter to the function of the SDDownloader class used for establishing connections to data sources. The SDDownloader class also supports adding additional routes to the routing table after a connection has been established. This kind of operation is required mostly when the data source is located in another subnet that differs from the subnet leased by the VPN gateway.

Currently, the SDDownloader class supports decoders written in Python, PHP, and Javascript programming languages. C and Fortran programs compiled statically are also supported. To call a certain decoder, one must describe how the decoder should be executed in the configuration file. The SDDownloader class also supports post-processing. The mechanism of executing a post-processing decoder is the same as the decoder execution. To control how a decoder runs, the SDDownloader class records the whole output produced by the decoder and its return code. All the recorded events are then passed to the RabbitMQ message broker so that the satellite operators could watch the execution process in real time.

The SDDownloader class uses the meta information from the instrument configuration files to access and load data into the database using the DBIface interface. But one can implement a customized way to load data by passing a program as a parameter to the SDDownloader class. The execution mechanism is the same as the decoder execution.

The SDDownloader class requires the storage to be mounted to the file system. By default, the SDDownloader class compresses data before moving them to the storage. All raw, level-0, and level-1 data are stored. Currently, this behavior can be changed only by editing the global configuration file of the satellite.

\section{Conclusion}
\label{conclusion}
A data processing, storage, and distribution system has been created using the SDDS framework and deployed at the Space Monitoring Data Center of Skobeltsyn Institue of Nuclear Physics, Lomonosov Moscow State University. The system is available at \url{https://downloader.sinp.msu.ru}. During eighteen months of operation, the system has been functioning correctly and showing its effectiveness and stability. Using the API provided by SDDS to add support for new satellites to the system is easy and requires a minimum of coding. In particular, adding support for the DSCOVR satellite and three satellites of the GOES-series has taken 14 hours and 12 hours accordingly. SDDS has been approved at SINP MSU as the base framework to be used in satellite service development in the future.

\section*{Acknowledgment}
\label{acknowledgment}
I would like to thank the Space Monitoring Data Center's team for collaboration in the development and support of RDB, metadata, and data access services. I would like to thank Dr. Vladimir Kalegaev for helpful discussions on satellite data processing and clear problem statements. Also, I would like to thank Dr. Alexander Kryukov for his support in all aspects. This project is supported by RSF grant \#16-17-00098.

\end{document}